\documentclass[aps,prc,twocolumn,floatfix,showpacs,preprintnumbers,amsmath,amssymb,nofootinbib,groupedaddress]{revtex4-1}
\usepackage{color}
\usepackage{graphicx}
\usepackage{dcolumn}    
\usepackage{multirow}
\usepackage{bm}         
\usepackage{sidecap}
\usepackage{hyperref}   
\usepackage{xcolor}
\usepackage{amsmath}
\usepackage{color, colortbl}

\usepackage{hyperref} 
\definecolor{dark-red}{rgb}{0.,0.,0}
\definecolor{dark-blue}{rgb}{0.,0.,1}
\definecolor{medium-blue}{rgb}{0,0,1}
\definecolor{gray}{rgb}{0.85,0.85,0.85}

\hypersetup{
    colorlinks, linkcolor={dark-red},
    citecolor={dark-blue}, urlcolor={medium-blue}
}

\begin{document}
%

\title{Multipole excitations in hot nuclei within the finite temperature quasiparticle random phase approximation framework}

\author{E. Y{\"{u}}ksel\textsuperscript{1}}
\email{eyuksel@yildiz.edu.tr}
\author{G. Col\`o\textsuperscript{2,3}}
\author{E. Khan\textsuperscript{4}}
\author{Y.F. Niu\textsuperscript{5}}
\author{K. Bozkurt\textsuperscript{1}}

\affiliation{\textsuperscript{1} Physics Department, Yildiz Technical University, 34220
Esenler, Istanbul, Turkey\\
             \textsuperscript{2} Dipartimento di Fisica, Universit\`a degli 
             Studi di Milano, via Celoria 16, I-20133 Milano, Italy\\	
							\textsuperscript{3} INFN, Sezione di Milano, Via Celoria 16, 20133 Milano, Italy \\				
               \textsuperscript{4} Institut de Physique Nucl\'eaire, Universit\'e Paris-Sud,
IN2P3-CNRS, F-91406 Orsay Cedex, France \\
               \textsuperscript{5} ELI-NP, Horia Hulubei National Institute for Physics and Nuclear Engineering, 30 Reactorului Street, RO-077125, Bucharest-Magurele, Romania}

\date{\today} 

%
\begin{abstract}
The effect of temperature on the evolution of the isovector dipole
and isoscalar quadrupole excitations in $^{68}$Ni and
$^{120}$Sn nuclei is studied within the fully self-consistent finite
temperature quasiparticle random phase approximation framework, based on the Skyrme-type SLy5 energy
density functional. The new low-energy excitations emerge due to the transitions from thermally occupied states to the discretized continuum at finite temperatures, whereas the isovector giant dipole resonance is not strongly impacted by the increase of temperature. The radiative dipole strength at low-energies is also investigated for the $^{122}$Sn nucleus, becoming compatible with the available experimental data when the temperature is included.
In addition, both the isoscalar giant quadrupole
resonance and low-energy quadrupole states are sensitive to the temperature effect: while the centroid energies decrease in the case of the isoscalar giant quadrupole resonance, the collectivity of the first $2^{+}$ state is quenched and the opening of new excitation channels fragments the low-energy strength at finite temperatures.
\end{abstract}

\pacs{21.60.Jz, 24.30.Cz}



\maketitle
\section{Introduction}
Giant resonances in nuclei take an important place in the field of nuclear physics, allowing us to shed light in a unique way on several aspects of the structure and properties of nuclei. 
For instance, dipole states are very sensitive to the isovector 
channel of the nucleon-nucleon effective interaction, and to the (a)symmetry part of
the nuclear equation of state. In the recent decades, the giant dipole resonances (GDRs) have been widely investigated and their features are well determined due to the combined use of the current theoretical models and experimental works. 
However, there are still some modes which are elusive and deserve some study.
In addition, investigation of the giant resonances under extreme conditions (neutron excess, temperature, etc.) provides complementary information about the structure and properties of nuclei and yet they constitute a further challenge for theory.

In recent years, it has been shown that multipole excitations of nuclei exhibit interesting features with increasing neutron number. For instance, in neutron-rich nuclei, the formation of low-energy dipole strength or pygmy dipole resonance (PDR) has been studied both with different theoretical models \cite{paar00,yuk12,roc12,lan09,vre01} and experimental techniques \cite{sav13,ozel14,wie09,krum15}. These results are quite important in order to understand the behavior of nuclei with the increase of the neutron number. In particular, the low-energy dipole states are related to the neutron skin thickness \cite{pie06}, symmetry energy \cite{klim07}, and also astrophysical processes \cite{gori02,lit09} like the $r$-process nucleosynthesis. 
Much less has been done regarding how these correlations evolve with temperature.
Since the nature of the low-energy modes is not clear yet, even at zero temperature, 
investigating their nature is generating a quite active field of research \cite{paar00,sav13,bra16}. 

The study of giant resonances in highly excited nuclei (hot nuclei) has also been a subject of interest due to their impact on  astrophysical processes. Experimentally, several groups tried to study the temperature dependence of the mean energy and width of the giant resonances. In the recent decades, giant dipole resonances built on excited states have been studied either using inelastic scattering of light particles or fusion-evaporation reactions \cite{ram96,bau98,kel99,kell99,heck03}. Results revealed that, while the GDR energy remains almost constant, this is not the case for the width, which changes both as a function of excitation energy and spin. A comprehensive review of recent studies about the effect of temperature on giant resonance properties is given in Refs. \cite{bor98,san06}.

Over the years, different theoretical models have been applied to study the effect of temperature on the evolution of the multipole response in nuclei: the random phase approximation (RPA) plus particle-vibration coupling (along the nuclear field theory 
formalism) \cite{bor86,don96}, the quasiparticle-phonon model \cite{st04}, 
the phonon damping model \cite{dan96,dan97}, and 
the extended time-dependent Hartree-Fock (TDHF) \cite{lac98,lac00}. 
While in most of these models there are phenomenological inputs,
the present work deals with the fully self-consistent quasiparticle random 
phase approximation (QRPA). In this way, although the 
damping width of the states under investigation can not be described, connections
between the properties of the states and the microscopic effective interaction can be made.
Long ago, the finite temperature RPA (FT-RPA) \cite{civi84,vaut84,bes84} with schematic interactions was introduced and used in order to investigate the properties of giant resonances with the temperature effect. In addition, the first self-consistent FT-RPA calculations were performed in Ref. \cite{saga84} in the case of the $^{40}$Ca nucleus. It was shown that the low-energy part of the isoscalar quadrupole response was rather sensitive to the temperature due to the particle-particle (pp) and hole-hole (hh) excitations which start to contribute at finite temperatures. However, the isovector dipole response of nuclei remained almost the same with increasing temperature. Recently, the multipole excitations of nuclei have also been studied using the self-consistent finite temperature relativistic RPA \cite{yifei09} and the emergence of low-energy strength has been predicted for monopole and dipole modes with increasing temperature.

The quasiparticle random phase approximation (QRPA) is known to be an appropriate and powerful tool for the description of the multipole excitations in open-shell nuclei. It allows for the study of the pairing effect in nuclear excitations, especially in the low-energy sector which is usually impacted. 
However, investigations of hot nuclei require the extension of the current models to the finite temperature QRPA (FT-QRPA), which is a rather complicated task. The finite temperature QRPA equations were first derived in Ref. \cite{som83} but the calculations were performed within a schematic model. Then, the finite temperature continuum-QRPA equations were derived using the Green's function formalism in Ref. \cite{khan04} to study the effect of temperature on the open-shell nuclei. However, calculations were not totally self-consistent because of the so-called Landau approximation for the velocity dependent terms of the residual Skyrme interaction.
The finite temperature continuum-QRPA formalism has also been used to study the low-energy enhancement of the radiative dipole strength ($\gamma$-ray strength) in medium-mass and heavy nuclei \cite{lit13}.
Thereafter, there is no further investigation available on the effect of temperature on the multipole response of the open-shell nuclei, neither with Skyrme or Gogny interactions nor within the relativistic framework. In particular, investigations within the fully self-consistent FT-QRPA are still lacking. 
Additional works are required in this field in order to elucidate the response of open-shell nuclei at finite temperatures with respect to the energy density functional framework.

The purpose of the present work is to investigate the effect of temperature on the multipole excitations of nuclei using for the first time the fully self-consistent FT-QRPA framework. The Skyrme-type energy density functional will be considered. The use of a microscopic self-consistent approach is important in order to describe nuclear structure under unusual conditions such as pairing and temperature coupled effects. 

The paper is organized as follows. In Sec. II, we briefly summarize the finite temperature mean-field theory and introduce the finite temperature QRPA formalism. In Sec. III, The FT-QRPA calculations are performed on top of the finite temperature Hartree-Fock BCS (FT-HFBCS) method in order to investigate the effect of temperature on the isovector dipole ($J^{\pi}=1^{-}$) and isoscalar quadrupole responses ($J^{\pi}=2^{+}$) in $^{68}$Ni and $^{120}$Sn nuclei. The radiative dipole strength function in the 
low-energy region is also investigated in $^{122}$Sn by increasing the temperature, and is compared with the available experimental data. Finally, summary and conclusions are given in Sec. IV.

\section{Microscopic Model: The Finite Temperature QRPA}
In this work, the finite temperature QRPA calculations have been carried out under the assumption of spherical symmetry, allowing for the calculation of giant resonances of various multipolarities in nuclei at
finite temperature. 

First, finite temperature HFBCS calculations were performed in coordinate space, yielding ground state properties of nuclei. At finite temperatures, the occupation probabilities of the states read
\begin{equation}
n_i=v_{i}^{2}(1-f_{i})+u_{i}^{2}f_{i},
\end{equation} 
where $u_i$ and $v_i$ are the BCS amplitudes. The temperature dependent Fermi-Dirac distribution function is given by
\begin{equation}
\label{eq:fd}
f_{i}=[1+exp(E_{i}/k_{B}T)]^{-1},
\end{equation}
where $E_{i}$ is quasiparticle (q.p.) energy, $k_{B}$ is the Boltzmann
constant, and $T$ is the temperature. Detailed information about the FT-HFBCS equations can be found in Refs. \cite{good81,yuk14}. 

In the present calculations, we use a zero-range density-dependent pairing interaction of surface type \cite{ber91},
\begin{equation}
V_{pair}(\textbf{r}_{1},\textbf{r}_{2})=V_{0}\left[1-\left(\frac{\rho(\textbf{r})}{\rho_{0}}\right)\right]\delta(\textbf{r}_{1}-\textbf{r}_{2}),
\end{equation}
where $\rho_{0}=$0.16 fm$^{-3}$ is the nuclear saturation density and 
$\rho(\textbf{r})$ is the particle density. The pairing strength $V_{0}$ is set for each nuclei according to the well-known three-point mass formula \cite{sat98}. The neutron pairing gap values are determined as $\Delta_{n}=1.6$ and $1.45$ MeV for $^{68}$Ni and $^{120}$Sn nuclei, respectively. In order to perform calculations at finite temperatures, the QRPA is extended to the finite temperature case.
Due to the effect of temperature, the excitation operator now involves 
both two-quasiparticle creation or annihilation operators (as in normal QRPA) and 
one-quasiparticle creation plus one-quasiparticle annihilation operators. This is due to the fact that at finite temperature the ground state is no longer the quasiparticle vacuum. The QRPA excitation operator reads
\begin{align}
\label{eq:ex}
\Gamma_{\nu}^{\dagger}=\sum_{a\geq b} \Big\{X_{ab}^{\nu}a_{a}^{\dagger}a_{b}^{\dagger}- Y_{ab}^{\nu}a_{b}a_{a} 
+P_{ab}^{\nu}a_{a}^{\dagger}a_{b} - Q_{ab}^{\nu}a_{b}^{\dagger}a_{a}\Big\}, 
\end{align}
where $a^{\dagger}$ and $a$ are the quasiparticle creation and destruction operators, respectively.
The finite temperature QRPA equations are obtained from the method of the equation of motion \cite{ring80,suho07}.
\allowdisplaybreaks
\begin{equation}
\label{eq:eom}
\left\langle BCS\right|\left[\delta\Gamma,[H,\Gamma_{\nu}^{\dagger}]\right]|BCS\rangle=E_{\nu}\left\langle BCS\right|\left[\delta\Gamma,\Gamma_{\nu}^{\dagger}\right]|BCS\rangle, 
\end{equation}
where $H$ is the Hamiltonian and $|\text{BCS}\rangle$ represents the thermal vacuum. The derivation of the expressions for the matrix elements in angular momentum coupled form is provided in the Appendix \ref{appendix}. The finite temperature QRPA matrix is given by
\begin{equation}
\left( { \begin{array}{cccc}\label{eq:qrpa}
 \widetilde{C} & \widetilde{a} & \widetilde{b} & \widetilde{D} \\
 \widetilde{a}^{+} & \widetilde{A} & \widetilde{B} & \widetilde{b}^{T} \\
-\widetilde{b}^{+} & -\widetilde{B}^{\ast} & -\widetilde{A}^{\ast}& -\widetilde{a}^{T}\\
-\widetilde{D}^{\ast} & -\widetilde{b}^{\ast} & -\widetilde{a}^{\ast} & -\widetilde{C}^{\ast}
 \end{array} } \right)
 \left( {\begin{array}{cc}
\widetilde{P}  \\
\widetilde{X }  \\
\widetilde{Y}  \\
\widetilde{Q} 
 \end{array} } \right)
 = E_{\nu}
  \left( {\begin{array}{cc}
\widetilde{P}  \\
\widetilde{X}  \\
\widetilde{Y}  \\
\widetilde{Q} 
\end{array} } \right), \end{equation}
where the eigenvectors and eigenenergies of the matrix are $\widetilde{P}, \widetilde{X}, \widetilde{Y}, \widetilde{Q}$, and $ E_{\nu}$, respectively. The FT-QRPA matrices are diagonalized in a self-consistent way,  providing a state-by-state analysis for each excitation. This is the main advantage of the present approach in configuration space.
The temperature dependencies of the matrices are given by
\begin{align}
\begin{split}
\widetilde{A}_{abcd}=&\sqrt{1-f_{a}-f_{b}} A'_{abcd}\sqrt{1-f_{c}-f_{d}}\\
&+(E_{a}+E_{b})\delta_{ac}\delta_{bd} \label{eq:temp},
\end{split}\\
\begin{split}
\widetilde{B}_{abcd}=&\sqrt{1-f_{a}-f_{b}} B_{abcd}\sqrt{1-f_{c}-f_{d}}, 
\end{split}\\
\begin{split}
\widetilde{C}_{abcd}=&\sqrt{f_{b}-f_{a}} C'_{abcd}\sqrt{f_{d}-f_{c}}\\
&+(E_{a}-E_{b})\delta_{ac}\delta_{bd},
\end{split}\\
\begin{split}
\widetilde{D}_{abcd}=&\sqrt{f_{b}-f_{a}} D_{abcd}\sqrt{f_{d}-f_{c}}, 
\end{split}\\
\begin{split}
\widetilde{a}_{abcd}=&\sqrt{f_{b}-f_{a}} a_{abcd}\sqrt{1-f_{c}-f_{d}}, 
\end{split}\\
\begin{split}
\widetilde{b}_{abcd}=&\sqrt{f_{b}-f_{a}} b_{abcd}\sqrt{1-f_{c}-f_{d}}, 
\end{split}\\
\begin{split}
\widetilde{a}_{abcd}^{+}=&\widetilde{a}_{abcd}^{T}=\sqrt{f_{d}-f_{c}} a_{abcd}^{+}\sqrt{1-f_{a}-f_{b}}, 
\end{split}\\
\begin{split}
\widetilde{b}_{abcd}^{T}=&\widetilde{b}_{abcd}^{+}=\sqrt{f_{d}-f_{c}} b_{abcd}^{T}\sqrt{1-f_{a}-f_{b}}.\label{eq:temp1}
\end{split}
\end{align}
where $E_{a(b)}$ is the quasiparticle energy of the states obtained from the FT-HFBCS results. In addition, the amplitudes read
\begin{gather}
\begin{aligned}
\widetilde{X}_{ab}=X_{ab}\sqrt{1-f_{a}-f_{b}},
\end{aligned} \\
\begin{aligned}
\widetilde{Y}_{ab}=Y_{ab}\sqrt{1-f_{a}-f_{b}},
\end{aligned} \\
\begin{aligned}
\widetilde{P}_{ab}=P_{ab}\sqrt{f_{b}-f_{a}},
\end{aligned}\\
\begin{aligned}
\widetilde{Q}_{ab}=Q_{ab}\sqrt{f_{b}-f_{a}}.
\end{aligned}
\end{gather}

It should be noted that the diagonal part of the FT-QRPA matrix includes both
($E_{a}+E_{b}$) and ($E_{a}-E_{b}$) configuration energies. In the FT-QRPA matrix the $\widetilde{A}$ and
$\widetilde{B}$ matrices describe the effects of the excitations of
quasiparticle pairs, which also survive at zero temperature.
The other components of the FT-QRPA matrix, $\widetilde{C}$, $\widetilde{D}$,
$\widetilde{a}$, $\widetilde{b}$, $\widetilde{a}^{+}$, and
$\widetilde{b}^{T}$ play a role with increasing temperature because they are impacted by the increasing changes in the occupation factors.

In the present work, the structure of the low-energy peaks is also analyzed using the FT-QRPA amplitudes.
For a given excited state $ E_{\nu}$, the contribution of the proton and neutron quasiparticle configurations to the excitation is determined by the FT-QRPA amplitudes
\begin{equation}
A_{ab}=|\widetilde{X}_{ab}^{\nu}|^{2}-|\widetilde{Y}_{ab}^{\nu}|^{2}+|\widetilde{P}_{ab}^{\nu}|^{2}-|\widetilde{Q}_{ab}^{\nu}|^{2},
\label{Aab}
\end{equation} 
and the normalization condition can be written as
\begin{equation}
\sum_{a\geq b}A_{ab}=1.
\label{aaa}
\end{equation}

At finite temperatures, the reduced transition probability for any operator $\hat{F}_{J}$ is given by

\begin{widetext}
\begin{equation}
\begin{split}
B(EJ,\widetilde0\rightarrow \nu)&=\bigl|\langle \nu ||\hat{F}_{J}||\widetilde0\rangle \bigr|^{2}\\
&=\biggl|\sum_{c\geq d}\Big\{(\widetilde{X}_{cd}^{\nu}+ \widetilde{Y}_{cd}^{\nu})(v_{c}u_{d}+u_{c}v_{d})\sqrt{1-f_{c}-f_{d}}+(\widetilde{P}_{cd}^{\nu}+\widetilde{Q}_{cd}^{\nu})(u_{c}u_{d}-v_{c}v_{d})\sqrt{f_{d}-f_{c}}\Big\}\langle c ||\hat{F}_{J}||d\rangle\biggr|^{2},
\end{split}
\label{bel}
\end{equation}
\end{widetext}
where $|\nu\rangle$  is the excited state and $|\widetilde0\rangle$ is the correlated FT-QRPA ground state.
In the present study, fully self-consistent calculations are performed;
namely, both the finite temperature HFBCS equations and the FT-QRPA
matrices are based on the same Skyrme energy density functional.
The Skyrme-type SLy5 interaction has been used in the calculations since it is well tailored for the description of the properties of exotic nuclei
\cite{sly5}. The continuum is discretized inside a spherical
box of 20 fm with 0.1 fm mesh size. We use a large
quasiparticle energy cut off ($E_{cut}=100$ MeV) allowing the energy
weighted sum rule (EWSR) to be satisfied: the maximum relative
difference with the theoretical EWSR is usually around 2.0\%.

It has been known that the phase transition of nuclei from superfluid to the normal state takes place at temperatures $T\approx 0.5-1$ MeV \cite{good81,yuk14} and the shape transition from deformed to spherical shape occurs above $T > 1$ MeV \cite{egi00}. Therefore, the deformation does not play a role above the critical temperatures, and use of the spherical FT-QRPA is relevant in the calculation of the multipole response of nuclei.
In addition, contributions from the continuum states become large at around $T\approx 4$ MeV \cite{bon84}. 
In order to avoid large contributions from the continuum states and unphysical neutron
vapor, calculations are performed up to $T=2$ MeV.

The Lorentzian averaging is known as a convenient tool to present the $E1$ and $E2$ strength in nuclei \cite{yuk12,roc12,lan09}. In the present work, the discrete spectrum is averaged with a
Lorentzian of $\Gamma=1$ MeV width using
\begin{equation}\label{lorentz}
S(EJ,E_{\nu})=\sum_{\nu}\frac{1}{2\pi}\frac{\Gamma}{(E-E_{\nu})^{2}-\Gamma^{2}/4}B(EJ,\widetilde0\rightarrow \nu),
\end{equation}
where $E_{\nu}$ is the excitation energy. In the present work, the centroid energy of the resonance is calculated with
\begin{equation}
E_{c}=\frac{m_{1}}{m_{0}},
\end{equation}
where the energy weighted moments $m_{1}$ and $m_{0}$ are defined using
\begin{equation}
m_{k}=\sum_{\nu}B(EJ,\widetilde0\rightarrow \nu)E_{\nu}^{k}.
\end{equation}

The limits of the FT-QRPA to the QRPA, RPA, and
FT-RPA cases have been checked: in the zero temperature limit, the
Fermi-Dirac distribution function goes to zero and the calculations
converge towards the QRPA, as expected. In the zero pairing limit, one obtains the
RPA and FT-RPA limits at zero and finite temperatures, respectively.

\section{results}
\subsection{Dipole Strength at Finite Temperatures} \label{a1}
We start by investigating the effect of temperature on the isovector dipole response of $^{68}$Ni and $^{120}$Sn nuclei. 
These two nuclei are benchmark in order to study the mass and neutron excess dependence on the
multipole excitations. In the finite temperature mean-field approach, nuclei undergo a sharp phase transition at critical temperatures $T_{c}$ due to the vanishing of the pairing correlations \cite{good81,yuk14}. 
In this work, the critical temperature values are calculated using the FT-HFBCS method: $T_{c}=$ 0.96 and 0.84 MeV for $^{68}$Ni and $^{120}$Sn nuclei, respectively. The value of the neutron pairing gap at zero temperature and the critical temperature generally follow the $T_{c}\approx0.57\Delta_{T=0}$ empirical rule, as expected \cite{good81,yuk14,khan07,niu13}.
It should be noted that the use of the grand-canonical description leads to sharp
phase transitions in nuclei within our model calculations. Recently, it has been shown that the use of the canonical description of nuclei at finite temperature removes sharp phase transitions in nuclei, and pairing correlations persist at high temperatures \cite{gam13}. However, the effect of the pairing correlations also weakens above the critical temperature. Therefore, our model is reliable to explore the qualitative changes in the multipole response in nuclei within our temperature range.

\subsubsection{$^{68}$Ni Nucleus}

In the neutron-rich $^{68}$Ni nucleus, the formation of the pygmy dipole resonance has been predicted within different theoretical models \cite{sav13,ter06,vre12,yuk12}. 
Recently, the pygmy dipole resonance was also obtained at
around 11 MeV using $\gamma$-decay, following Coulomb excitation of the nucleus on a gold target
\cite{wie09}. Later, another Coulomb excitation experiment in inverse kinematics was performed on $^{68}$Ni, 
and the GDR and pygmy dipole resonance energies were obtained at 17.1(2) and 9.55(17) MeV, respectively \cite{ros13}. Since the pygmy dipole strength has already been observed in $^{68}$Ni nucleus, it would also be interesting to investigate the effect of temperature on this region. 

Before discussing the effect of temperature on the dipole response of $^{68}$Ni nucleus, explaining its main effect on the proton and neutron states is also necessary to understand the underlying mechanism driven by the temperature. With increasing temperature, nucleons are promoted to higher energy states, which eventually increases (decreases) the occupation probabilities of states above (below) the Fermi level. For instance, in the $^{68}$Ni nucleus, neutron $1f_{5/2}$ state below the Fermi level and $1g_{9/2}$ state above the Fermi level are partially occupied due to pairing effects at zero temperature. By increasing the temperature, neutrons are mainly promoted from $2p_{3/2}$, $2p_{1/2}$, and $1f_{5/2}$ states below the Fermi level to $1g_{9/2}$, $2d_{5/2}$, and $3s_{1/2}$ states above the Fermi level. The effect of temperature is also similar for proton states.
At zero temperature, proton states are fully occupied up to $1f_{7/2}$ state and form $Z$=28 shell closure, as expected. By increasing the temperature, protons are also promoted from $1d_{3/2}$, $2s_{1/2}$, and $1f_{7/2}$ states to $2p_{3/2}$, $1f_{5/2}$, $2p_{1/2}$, and $1g_{9/2}$ states. Therefore, new excitation channels become possible due to the thermally unblocked states at finite temperature.

\begin{figure}[!ht]
  \begin{center}
\includegraphics[width=1\linewidth,clip=true]{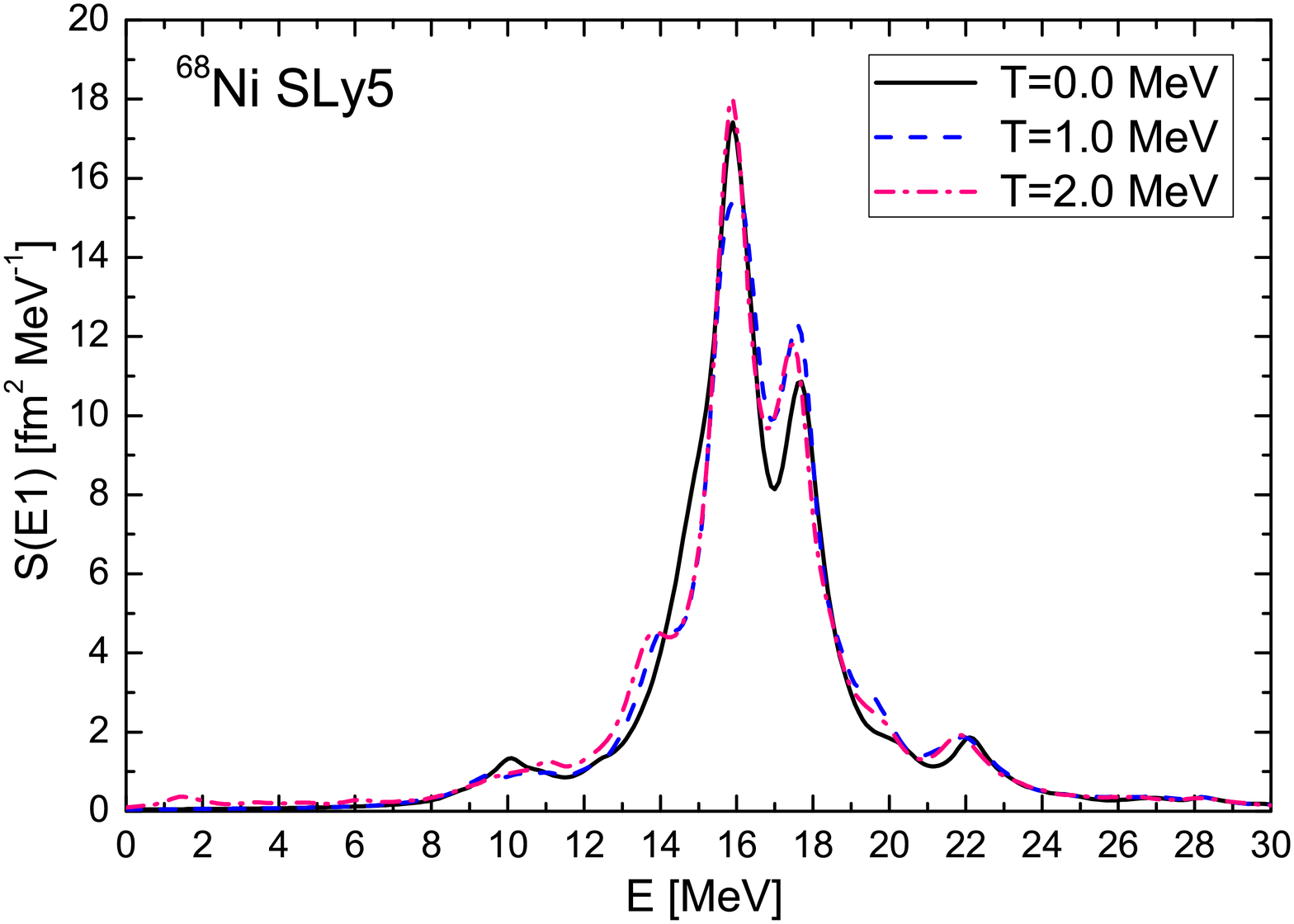}
\includegraphics[width=1\linewidth,clip=true]{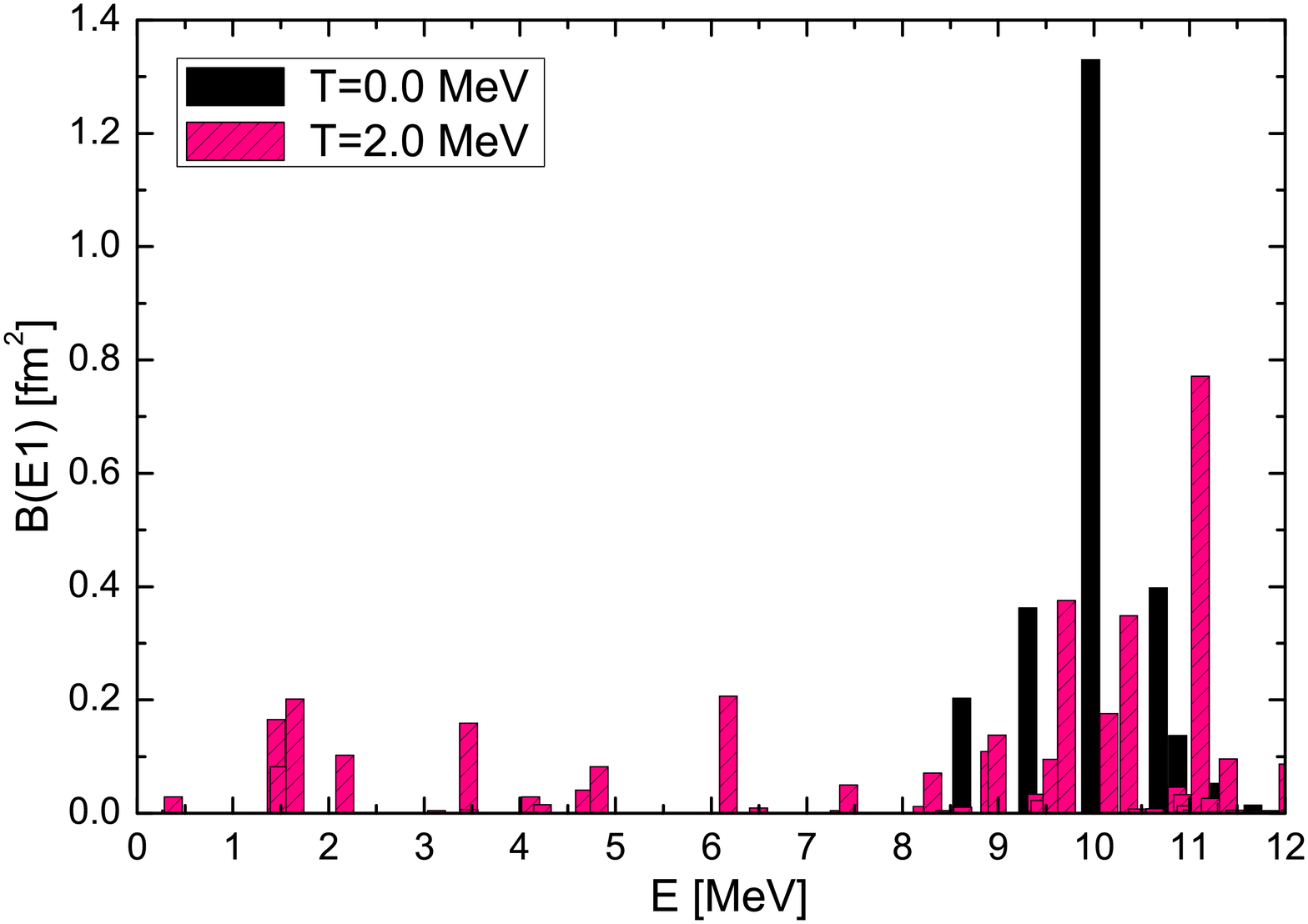}
  \end{center}
 \caption{Upper panel: The isovector dipole strength function in $^{68}$Ni calculated with FT-QRPA and the Skyrme-type SLy5 interaction at 
	$T=0$, $1$, and $2$ MeV. Lower panel: The reduced transition probabilities for the low-energy dipole region at $T=0$ MeV and $T=2$ MeV.} 
  \label{681-}
\end{figure}

In the upper panel of Fig. \ref{681-} we present the isovector dipole strength function in $^{68}$Ni nucleus at $T=0$, $1$, and $2$ MeV, respectively. At zero temperature, the centroid energy of the well known isovector giant dipole resonance (IVGDR) is calculated between 0 and 30 MeV and is obtained at 16.8 MeV. In addition, the pygmy dipole strength is found around E $\approx$ 10 MeV and exhausts $1.2\%$ of the EWSR with the present interaction. Our results are in good agreement with recent experimental results \cite{ros13} and theoretical works \cite{yuk12,roc12,ter06,vre12}.
The IVGDR strength and energy slightly change with increasing temperature. For instance, at $T=2$ MeV, the centroid energy is found at 16.5 MeV. The effect of temperature on the dipole response function is more striking on the low-energy part. 
The new excited states appear below 8 MeV, as seen in the lower panel of Fig. \ref{681-}. In order to understand the origin of these new excited states, the quasiparticle configurations are analyzed in Table \ref{table:xx} at $T=2$ MeV. 
In particular, these states are not collective and are made of almost single neutron transitions from thermally unblocked and loosely bound neutron $2d_{5/2}$ and $3s_{1/2}$ states to the continuum at finite temperature. Since the continuum is discretized, it should be mentioned that the continuum effects are not taken into account exactly in our model. Therefore, these transitions in the low-energy region are just representing the transitions to the single-particle states in the discretized continuum, and an important effect can be expected with the exact treatment of the continuum. Nevertheless, the results show that the continuum plays an essential role in the formation of these new low-energy excited states at finite temperatures. This important effect of the continuum will also be analyzed in Sec. \ref{rds} by comparing with the experimental data in the case of the $\gamma$-ray strength function for $^{122}$Sn.

\begin{table}[ht]
\caption{The selected low-energy excitations for the $^{68}$Ni nucleus at $T=2$ MeV. The excited state energies, configurations, and their contributions (in percentage) to the norm of the states [see Eqs. (\ref{Aab}) and (\ref{aaa})] are presented, respectively. Herein, $\pi$ and $\nu$ refer to the proton and neutron, respectively.} 
\centering 
\begin{tabular}{c c c c} 
\hline\hline  \\[-1.0em]
 Energy & Configuration & \% \\ [1ex] 
\hline \\[-1.0em] 
$E=1.36$ MeV & $\nu3s_{1/2}\rightarrow \nu3p_{3/2} $ & 99.8 \\ 
$E=1.39$ MeV & $\nu3s_{1/2}\rightarrow \nu3p_{1/2} $ & 99.8 \\ 
$E=2.07$ MeV & $\nu2d_{5/2} \rightarrow \nu3p_{3/2} $ & 99.8 \\ 
$E=3.37$ MeV & $\nu2d_{5/2}\rightarrow \nu2f_{7/2}$ & 99.9 \\ 
$E=4.73$ MeV  & $\nu2d_{5/2}\rightarrow \nu4p_{3/2}$  & 99.9 \\
$E=6.08$ MeV &  $\nu2d_{5/2}\rightarrow \nu3f_{7/2}$& 99.7  \\
\hline \\ [-1.ex]
\end{tabular}
\label{table:xx} 
\end{table}
\begin{table}[ht]
\caption{The major low-energy dipole excitations for $^{68}$Ni at $T=0$ and $2$ MeV. The configurations and their contributions to the norm of the states (in percentage) are displayed for each excitation energies, separately. The transitions that appear by increasing the temperature are also shown in bold.} 
\centering 
\begin{tabular}{c c c c c} 
\hline\hline  \\[-1.0em]
 Configuration & $T=0$ MeV & $T=2$ MeV & $T=2$ MeV\\ [1ex] 
 &E=10.05 MeV & E=9.62 MeV & E=11.0 MeV \\
\hline \\[-1.0em] 
$\nu1f_{5/2} \rightarrow \nu2d_{3/2}$ & 61.4 & 53.0 &-\\ 
$\nu1f_{7/2} \rightarrow \nu1g_{9/2} $  & 9.7 & 4.6& 9.4\\
$\nu2p_{3/2}\rightarrow \nu2d_{5/2} $& 7.1 & 5.5 & 4.8\\
$\nu1f_{5/2}\rightarrow \nu2d_{5/2}$ & 5.0 & 2.4 & - \\ 
$\nu1f_{5/2}\rightarrow \nu3d_{3/2} $ & 2.6 & 1.2 & - \\ 
$\nu1g_{9/2} \rightarrow \nu2h_{11/2}$ & - & \textbf{2.6} &\textbf{18.7}  \\ 
$\nu2d_{5/2}\rightarrow \nu4f_{7/2} $ & - & \textbf{5.9}& - \\
$\nu2p_{3/2} \rightarrow \nu3s_{1/2} $ & - & - & \textbf{4.8} \\ 
$\pi1f_{7/2}\rightarrow \pi1g_{9/2} $  & 4.4 & 1.9& 2.0 \\
$\pi1f_{5/2} \rightarrow \pi2d_{5/2}$ & - & \textbf{10.1} & - \\ 
$\pi2p_{3/2} \rightarrow \pi2d_{5/2}$ & - & \textbf{1.4} & \textbf{47.0} \\ 
$\pi1d_{5/2}\rightarrow \pi1f_{7/2}$ & - &  & \textbf{3.5} \\ 
$\pi2s_{1/2}\rightarrow \pi2p_{3/2}$ & - & - & \textbf{2.2} \\ 
\hline \\ [-1.ex]
\end{tabular}
\label{table:xy} 
\end{table}

The second important change caused by temperature is the fragmentation of the pygmy dipole states. At $T=2$ MeV, the main PDR state at 10 MeV is fragmented into several states, and another state, less strong than the one at $T=0$ MeV, appears at E=11.0 MeV.
The configurations for these excitations in the pygmy dipole region are given in Table \ref{table:xy} at $T=0$ MeV and $T=2$ MeV.
The region of the pygmy dipole resonance displays some collectivity at $T=0$ MeV, but this decreases at T=2 MeV.
At $T=0$ MeV and E=10.05 MeV, the main contributions come from the neutron quasiparticle excitations, as expected.
At $T=2$ MeV, the pygmy dipole resonance region is fragmented into several states, and two main low-energy peaks are found at 9.62 MeV and 10.3 MeV, exhausting $0.32\%$ and $0.31\%$ of the EWSR, respectively. 
Comparing the PDR states at E=10.05 MeV ($T=0$ MeV) and E=9.62 MeV ($T=2$ MeV), temperature does not impact configurations. However, some new excitation channels (displayed in bold in Table \ref{table:xy}) are opened and contribute to the excited states 
at $T=2$ MeV. The configuration analysis of these states shows that in addition to the neutron states, thermally occupied proton states also lead to the formation of the new excitation channels
and start to contribute to the PDR region with increasing
the temperature. Especially, thermally unblocked proton states play a major role at E=11.0 MeV at $T=2$ MeV (see Table \ref{table:xy}).

Let us now examine the contributions of new quasiparticle excitations to the dipole strength function. As we mentioned before,
the diagonal part of the FT-QRPA matrix takes contributions from ($E_{a}-E_{b}$) in addition to ($E_{a}+E_{b}$) two-q.p. energies [see Eq. \ref{eq:qrpa}]. While the IVGDR region mainly takes contributions from ($E_{a}+E_{b}$) two-q.p. energies and is less sensitive to the temperature effects due to its collective nature, contributions of the ($E_{a}-E_{b}$) two-q.p. energies mainly impact the low-energy region of the dipole spectrum and lead to an increase of the low-energy dipole strength due to the transitions from thermally unblocked states to the discretized continuum.

\subsubsection{$^{120}$Sn Nucleus}
The tin isotopes have also been the subject of many theoretical and experimental studies over
the years. The formation of low-energy dipole strength in the
$^{120}$Sn nucleus has been predicted in several theoretical works \cite{ter06,sar04} and experimental data is also available \cite{ozel14,krum15}. The occurrence of the low-energy dipole states has been observed in $^{120}$Sn using a $(\gamma,\gamma')$ experiment below the neutron emission threshold \cite{ozel14}. In this experiment, low-energy dipole states were obtained between 4-9 MeV and exhaust 0.22\% of the EWSR. In a more comprehensive experiment, the electric dipole strength in $^{120}$Sn nucleus was investigated with proton inelastic scattering below and above the threshold \cite{krum15}. While the centroid of the GDR is found at 15.0 MeV, several low-energy dipole excitations have been measured between 4 and 9 MeV. This low-energy area is peaked at 8.3 MeV and exhausts 2.3(2)\% of the EWSR.

\begin{figure}[!ht]
  \begin{center}
\includegraphics[width=1\linewidth,clip=true]{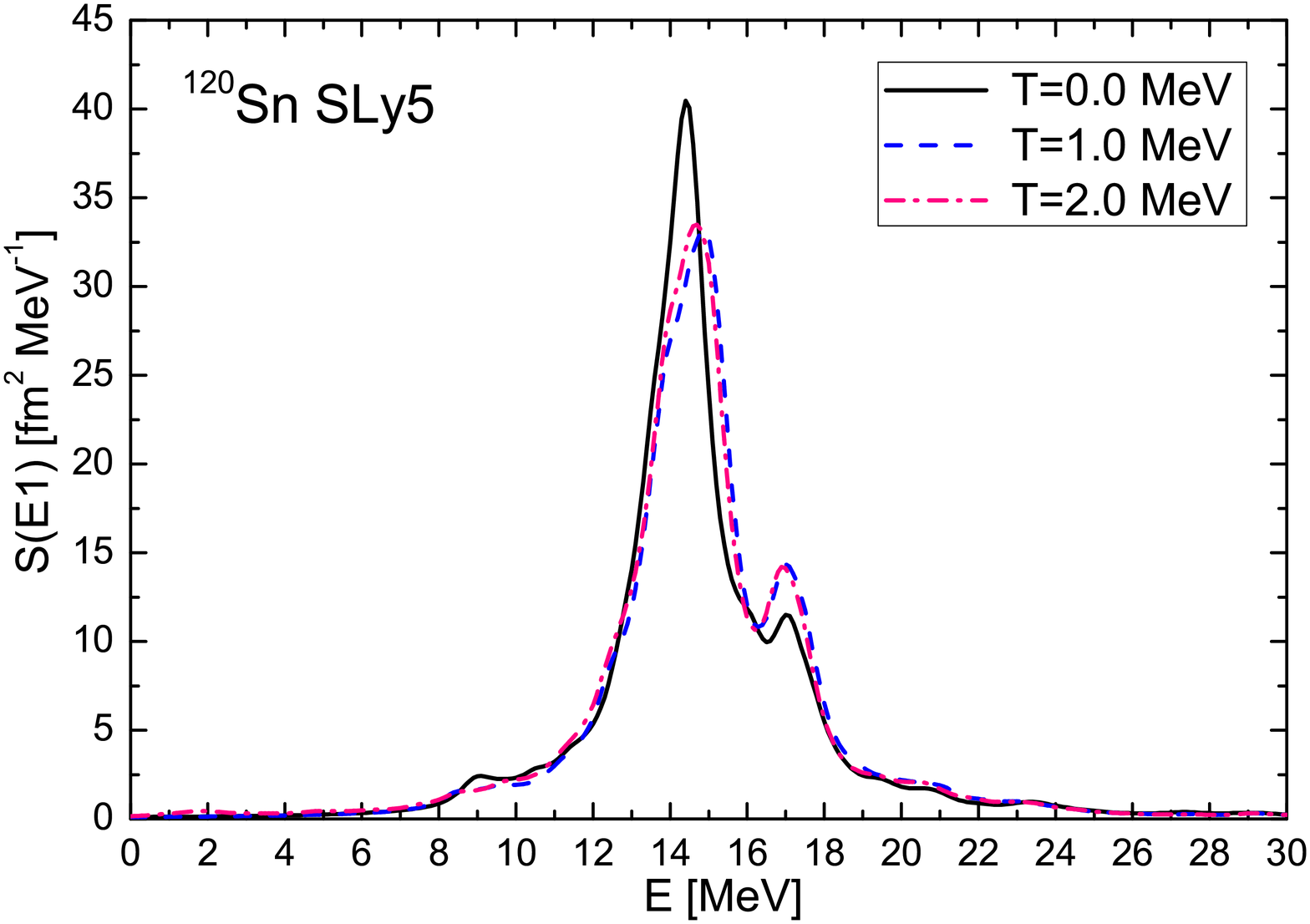}
\includegraphics[width=1\linewidth,clip=true]{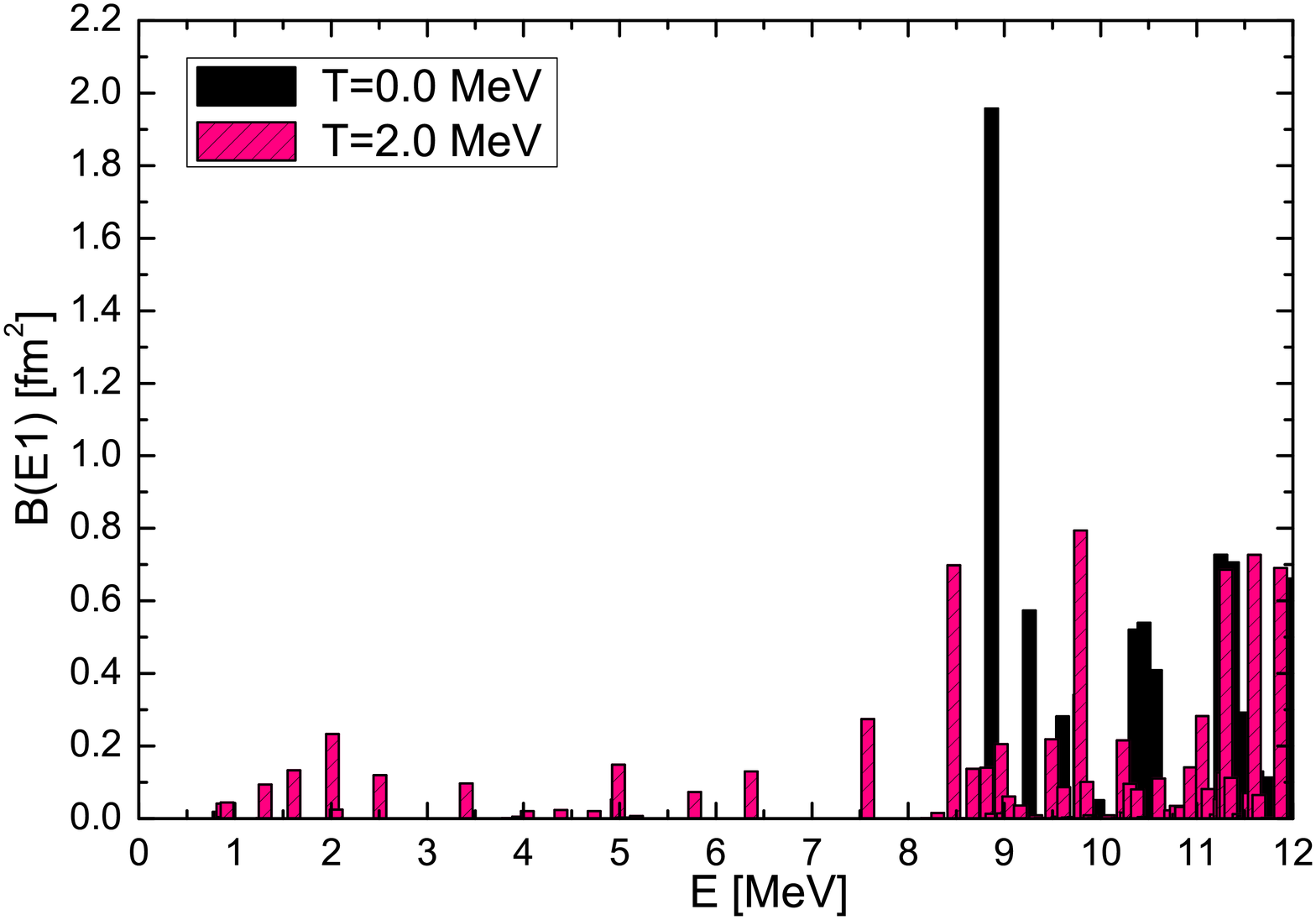}
  \end{center}
  \caption{Same as Fig. \ref{681-} but for the $^{120}$Sn nucleus.} 
  \label{1201-}
\end{figure}

The effect of temperature
on the evolution of the isovector dipole strength in
$^{120}$Sn is displayed in Fig. \ref{1201-}. At zero temperature, the
IVGDR centroid energy between 0 and 30 MeV is obtained at 15.0 MeV. In addition,
the first low-energy strength is found at 8.9 MeV
and exhausts 0.88\% of the EWSR. Our results are comparable with the recent experimental result of Ref. \cite{krum15} and the theoretical studies of Refs. \cite{vre01,sar04,ter06}.
The low-energy dipole state is mainly associated with neutron quasiparticle excitations. We find that
the major configurations at E=8.9 MeV are: proton $1g_{9/2}\rightarrow 1h_{11/2}$ (12.5\%) and neutron $1g_{7/2}\rightarrow 2f_{7/2}$ (33.6\%), $1g_{9/2}\rightarrow 1h_{11/2}$ (11.0\%), and $3s_{1/2}\rightarrow 3p_{3/2}$ (8.4\%). Therefore, this state displays some collectivity.

With increasing temperature, neutrons are mainly promoted from $2d_{5/2}$, $1g_{7/2}$, $3s_{1/2}$, and $2d_{3/2}$ states below the Fermi level into $1h_{11/2}$, $2f_{7/2}$, and $3p_{3/2}$ states above the Fermi level. While $^{120}$Sn nucleus is fully occupied up to proton $1g_{9/2}$ state and has a shell closure at Z=50, protons are also promoted from $2p_{3/2}$, $2p_{1/2}$, and $1g_{9/2}$ states
to $2d_{5/2}$, $1g_{7/2}$, $2d_{3/2}$, and $3s_{1/2}$ states with increasing temperature.
In the $^{120}$Sn nucleus, the evolution of the isovector dipole spectrum as a function of the temperature is quite similar to the one of the $^{68}$Ni nucleus. In the GDR region, the strength and centroid energy slightly decrease with increasing temperature. For instance, the centroid energy for the IVGDR is obtained at 14.95 MeV at $T=2$ MeV. 
We obtain an enhancement in the dipole strength at around E $\approx$ 17 MeV with the contribution of new quasiparticle excitations at finite temperatures. While the transitions from thermally unblocked and loosely bound neutron $2f_{7/2}$ and $3p_{3/2}$ states to the discretized continuum lead to the formation of the new non-collective excited states in the very low-energy region, the pygmy dipole resonance region seems to be fragmented into several states with comparable strengths at finite temperatures.
Below 10 MeV, we obtain two important peaks at $E=8.4$ MeV and $E=9.8$ MeV which exhaust 0.29\% and 0.38\% of the EWSR at $T=2$ MeV. 
Similarly to the $T=0$ MeV case, these states have some collectivity, albeit reduced with respect to $T=0$ MeV.
While the excitations are mainly neutron dominated at $E=8.4$ MeV, proton excitations gain importance at $E=9.8$ MeV with the contribution of thermally populated proton states above Z=50 shell. We find two major configurations at $E=9.8$ MeV: proton $2d_{5/2}\rightarrow 2f_{7/2}$ (50.2\%) and neutron $2d_{3/2}\rightarrow 2f_{5/2}$ (34.8\%). Therefore, the low-energy dipole strength also takes important contributions from thermally occupied proton states at finite temperatures.

\subsection{The Low-Energy Enhancement of the Radiative Dipole Strength: $^{122}$Sn Nucleus} \label{rds}
\begin{figure}[!ht]
  \begin{center}
\includegraphics[width=1\linewidth,clip=true]{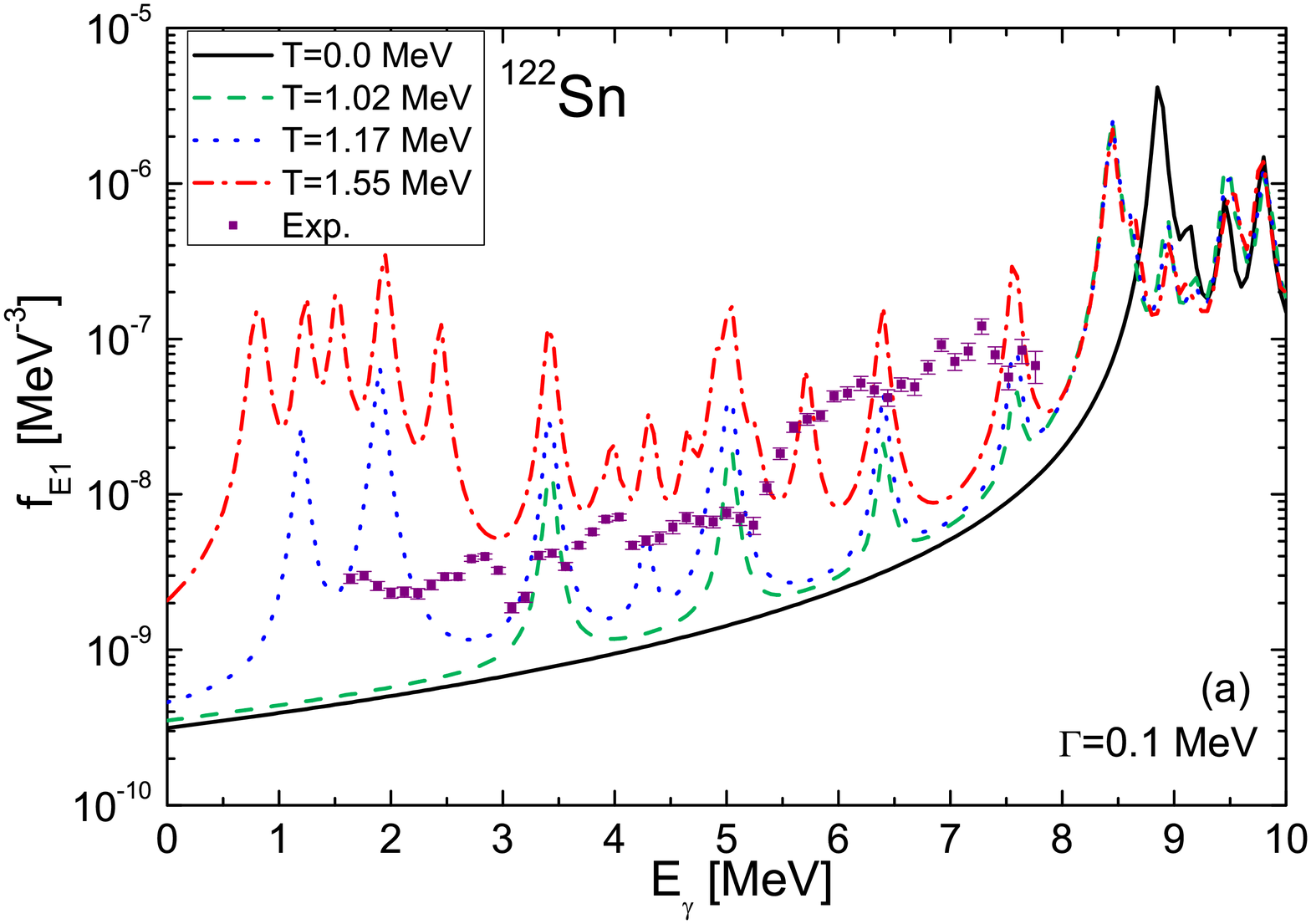}
\includegraphics[width=1\linewidth,clip=true]{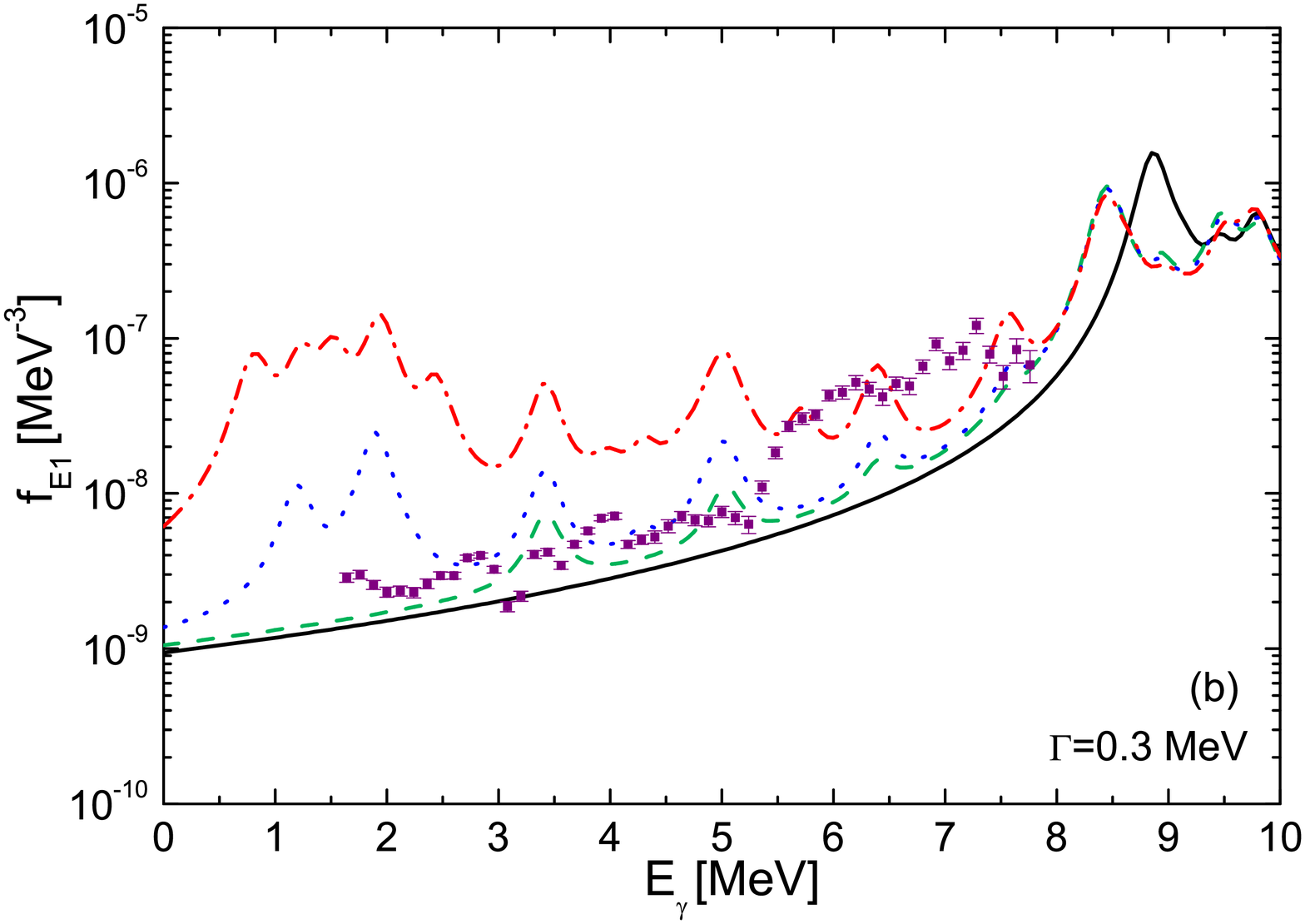}
  \end{center}
  \caption{The $\gamma$-ray strength function in $^{122}$Sn at $T=0$, $1.02$, $1.17$, and $1.55$ MeV using the Skyrme-type SLy5 interaction. The discrete spectrum is broadened by using Lorentzian functions with 0.1 MeV (a) and 0.3 MeV (b) width. The experimental data are taken from Ref. \cite{tof11}.}
	\label{122}
\end{figure}

Recently, an enhancement of the dipole strength at very low energies ($E\leq3-4$ MeV) has been obtained in several experimental studies of the radiative dipole strength \cite{gut05,lar10, tof10, tof11}. 
In Ref. \cite{lit13}, the authors performed finite temperature continuum QRPA and showed that transitions from thermally occupied states to the continuum can explain the observed low-energy enhancement of the radiative dipole strength. 
In order to check for a possible enhancement of the dipole strength at very low-energies, we compare our predictions with the results of Refs. \cite{lit13,tof11}, where the temperature range for $^{122}$Sn nucleus is taken as 1.02$\leq T \leq$1.17 MeV. Exploration of the radiative dipole strength at $T=$1.55 MeV is also shown in order to qualitatively investigate the effect of high temperature in the low-energy region. On this purpose, we performed FT-QRPA calculations in $^{122}$Sn at $T=0$, $1.02$, $1.17$, and $1.55$ MeV, displaying the $\gamma$-ray strength function ($f_{E1}$) in Fig.\ref{122}.
The $\gamma$-ray strength function and the dipole strength function [see Eq. \ref{lorentz}] are related to each other via \cite{lit09}
\begin{equation}
f_{E1}(E_{\gamma})=\frac{16 \pi e^{2}}{27(\hbar c)^{3}}S(E1,E_{\gamma}).
\end{equation}

In Fig.\ref{122}(a), the discrete dipole spectrum is broadened by using a Lorentzian function having 0.1 MeV width and the $\gamma$-ray strength function is displayed between 0 and 10 MeV, below the neutron separation energy. The main physical prediction is the increase of the low-energy strength with temperature, due to the opening of the new excitation channels and allowing for a better description of data. The QRPA results at zero temperature do not predict an enhancement of the dipole strength at very low energies. At $T=1.02$ MeV, we start to obtain an increase in the $\gamma$-ray strength function above E$>$3 MeV with the contribution of new excitation channels. By further increasing the temperature, the quasiparticle energies do not change drastically and the location of some excitations remains almost same. However, new excitation channels emerge and the dipole strength also increases due to the increasing diffuseness of the Fermi surface at higher temperatures. Above E$>$3 MeV, our results are compatible with the available experimental data by assuming temperatures at $T=1.02$ and $1.17$ MeV. Although we obtain new excited states below E$<$3 MeV, the FT-QRPA overestimates the $\gamma$-ray strength function in this region for $T\geq1.17$ MeV. Since the continuum is not taken into account exactly in our model, it is also possible to obtain a broad strength function in the low-energy region by changing the smoothing parameter in the Lorentzian function. In order to check the impact of the smoothing parameter in the low-energy region, the $\gamma$-ray strength function with a 0.3 MeV width is also displayed in Fig.\ref{122}(b). Although the effect of the smoothing parameter is non-negligible, it does not influence the qualitative discussion about the low-lying strength that we carry out in this Section.

\begin{table}[ht]
\caption{Same as in Table \ref{table:xx} but for $^{122}$Sn at $T=1.55$ MeV.} 
\centering 
\begin{tabular}{c c c c} 
\hline\hline  \\[-1.0em]
 Energy & Configurations & \% \\ [1ex] 
\hline \\[-1.0em] 
$E=1.24$ MeV & $\nu3p_{3/2}\rightarrow \nu4s_{1/2} $ & 99.6 \\ 
$E=1.94$ MeV & $\nu3p_{3/2}\rightarrow \nu3d_{5/2} $ & 99.9 \\ 
$E=3.42$ MeV & $\nu2f_{7/2} \rightarrow \nu3d_{5/2} $ & 99.9 \\ 
$E=4.30$ MeV & $\nu3p_{3/2} \rightarrow \nu5s_{1/2} $ & 99.9 \\ 
$E=4.91$ MeV & $\nu3p_{3/2} \rightarrow \nu4d_{5/2} $ & 99.9 \\ 
$E=5.03$ MeV & $\nu2f_{7/2}\rightarrow \nu2g_{9/2}$ & 99.9 \\ 
$E=6.40$ MeV  & $\nu2f_{7/2}\rightarrow \nu4d_{5/2}$  & 99.9 \\
$E=7.57$ MeV &  $\nu2f_{7/2}\rightarrow \nu3g_{9/2}$& 99.7  \\
\hline \\ [-1.ex]
\end{tabular}
\label{table:xq} 
\end{table}

In order to analyze this increase in the low-energy strength, the configurations for the selected low-energy excited states are also displayed in Table \ref{table:xq}. As mentioned before, the continuum plays an essential role in the formation of the very low-energy strength at finite temperatures. The transitions from thermally unblocked neutron $2f_{7/2}$ and $3p_{3/2}$ states to the discretized continuum create the low-energy $\gamma$-ray strength.
This kind of transition does not exist in the low-energy region of the dipole response at zero temperature but appears with increasing temperature. This increased dipole strength at finite temperature is important for the astrophysical events \cite{gori02,lit09}. For instance, the very low-energy dipole strength may have an impact on the neutron capture rates for $r$-process nucleosynthesis. The FT-QRPA provides a description of the underlying mechanism of these very low-energy transitions.
Our results also confirm the theoretical interpretations given in Ref. \cite{lit13}.
It should be noted that a proper treatment of the low-energy excitations necessitates more complete microscopic models. For instance,
treatment of nuclei in a canonical ensemble may reduce the very low-energy $\gamma$-ray strength and give a better agreement with the experimental data. While the traditional Q(RPA) does not predict any strength at low energies, inclusion of more complex configurations like 2p-2h also predicts new excited states in the low-energy region \cite{gam1,gam11}. However, these kinds of microscopic models do not exist at finite temperatures. The present FT-QRPA can be used as a second starting point to develop more complete microscopic models at finite temperatures.

\subsection{Quadrupole Strength at Finite Temperatures}
\subsubsection{$^{68}$Ni Nucleus}
\begin{figure}[!ht]
  \begin{center}
\includegraphics[width=1\linewidth,clip=true]{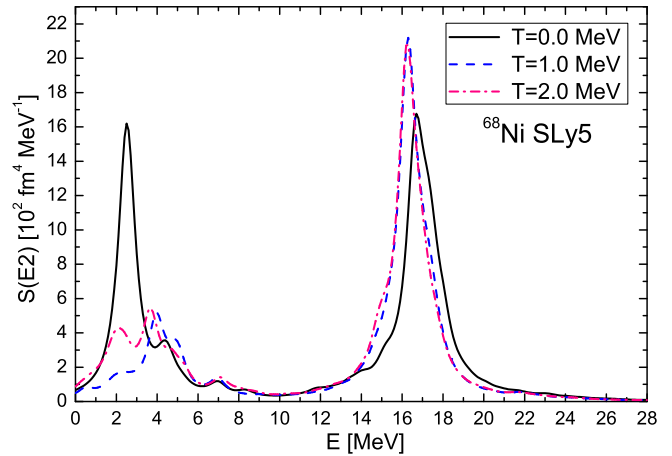}
  \end{center}
  \caption{The isoscalar quadrupole strength function in $^{68}$Ni calculated with the FT-QRPA and the Skyrme-type SLy5 interaction at 
	$T=0$, $1$, and $2$ MeV.}
  \label{682+}
\end{figure}

In this subsection, we analyze the effect of temperature on the isoscalar quadrupole response of nuclei. 
Recently, the isoscalar giant quadrupole resonance (ISGQR) centroid energy of $^{68}$Ni was determined at $15.9\pm1.3$ MeV using inelastic alpha and deuteron scattering \cite{van15}, and the first $2^{+}$ state was found at 2.034 MeV \cite{ra01}. In Fig. \ref{682+}, the isoscalar quadrupole strength function is displayed for $^{68}$Ni at finite temperatures.
In our calculations at zero temperature, the ISGQR and the first $2^{+}$ state are rather collective and located at 16.8 and 2.5 MeV, respectively. The major configurations for the first $2^{+}$ state are proton $1f_{7/2}\rightarrow 2p_{3/2}$ (15.9\%) and neutron $1f_{5/2}\rightarrow 1f_{5/2}$ (31.5\%), $2p_{1/2}\rightarrow 1f_{5/2}$ (20.6\%), and $1g_{9/2}\rightarrow 1g_{9/2}$ (17.2\%).

\begin{table}[ht]
\caption{Same as in Table \ref{table:xx} but for the low-energy quadrupole excitations in $^{68}$Ni at $T=2$ MeV. The transitions from thermally unblocked proton states are also shown in bold.} 
\centering

\begin{tabular}{c c c c} 
\hline\hline  \\[-1.0em]
 Energy & Configurations & \% \\ [1ex] 
\hline \\[-1.0em] 
$E=1.54$ MeV & $\pi2p_{3/2} \rightarrow \pi1f_{5/2}$ & \textbf{98.0}\\
$E=1.92$ MeV  & $\pi2p_{3/2}\rightarrow \pi2p_{1/2}$  &  \textbf{72.5} \\
            & $\nu2p_{3/2}\rightarrow \nu2p_{1/2}$& 19.1  \\
						& $\nu2d_{5/2}\rightarrow \nu2d_{3/2}$& 4.1  \\
$E=2.12$ MeV  & $\nu3s_{1/2}\rightarrow \nu3d_{5/2}$ &99.8  \\
$E=2.18$ MeV & $\pi2p_{3/2}\rightarrow \pi2p_{1/2}$  &  \textbf{20.8}  \\ 
							& $\nu2p_{3/2}\rightarrow \nu2p_{1/2}$  & 77.7  \\ 
$E=2.45$ MeV & $\nu2p_{3/2}\rightarrow \nu1f_{5/2} $  & 96.9  \\ 
$E=3.66$ MeV  & $\pi1f_{7/2}\rightarrow \pi2p_{3/2}$  &  \textbf{10.0}  \\ 
           & $\nu1f_{7/2}\rightarrow \nu2p_{3/2} $& 2.6  \\
           & $\nu1g_{9/2}\rightarrow \nu2d_{5/2} $& 81.2 \\
\hline \\ [-1.ex]
\end{tabular}
\label{table:zz} 
\end{table}

The effect of temperature is more striking on the ISGQR than on the IVGDR.
The temperature impacts both the ISGQR and low-energy
regions. First, the ISGQR
strength increases and the centroid energy decreases with increasing temperature.
For instance, the ISGQR centroid energies between
10 and 22 MeV are found at 16.8, 16.4, and
16.3 MeV at $T=0$, $1$ and $2$ MeV, respectively. Second, the first $2^{+}$ state strength is quenched already at $T=1$ MeV. It has been known that pairing correlations play a crucial
role in the isoscalar quadrupole
response in nuclei. Especially, the low-energy region is rather sensitive to the pairing effects \cite{scam13}. 
The main reason for these changes in the isoscalar quadrupole response is the disappearance of the pairing correlations after the critical temperature.
Furthermore, the isoscalar quadrupole strength starts to increase at
around $E\approx4$ MeV due to the smearing of the Fermi
surface and the contribution of the new excitation channels
at $T=1$ MeV. By further increasing the temperature, the isoscalar quadrupole strength also increases below $E<4$ MeV. This is due to the increasing diffuseness of the Fermi surface at higher temperatures. In order to clarify this enhancement in the low-energy strength, the major transitions contributing to this region are listed in Table \ref{table:zz} for $^{68}$Ni at $T=2$ MeV.
By increasing temperature, the low-energy quadrupole strength is impacted due to the changing occupation probabilities of states and mainly takes contributions from the excitations below and above the Fermi level. For instance, the smearing of the Fermi surface opens new excitation channels for the $2^{+}$ states within the proton $p$ -$f$ shell with increasing temperature, which eventually contribute to the low-energy region (shown in bold in Table \ref{table:zz}). The excitations from thermally occupied states to the discretized continuum are also obtained, whereas their contribution to the low-energy strength is low compared to excitations around Fermi level.

\subsubsection{$^{120}$Sn Nucleus}
\begin{figure}[!ht]
  \begin{center}
\includegraphics[width=1\linewidth,clip=true]{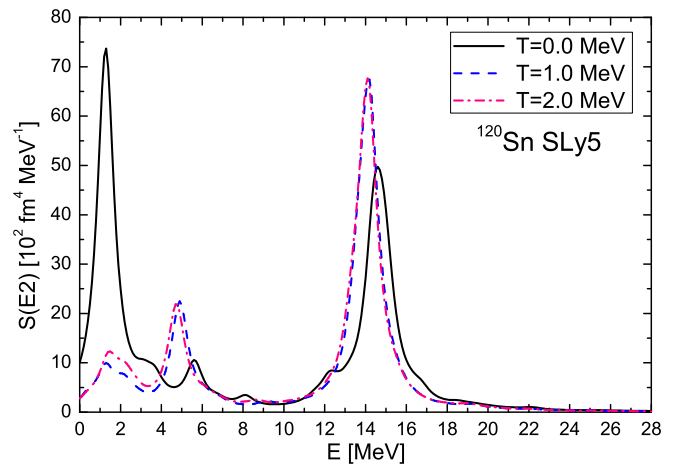}
  \end{center}
  \caption{Same as in Fig. \ref{682+} but for the $^{120}$Sn nucleus.} 
  \label{1202+}
\end{figure}

In Figure \ref{1202+}, the isoscalar quadrupole strength function is displayed for the $^{120}$Sn nucleus at $T=0$, $1$, and $2$ MeV. 
Experimentally, the ISGQR and the first $2^{+}$ state energies in the $^{120}$Sn nucleus were measured at $12.9\pm0.1$ MeV \cite{li10} and 1.17 MeV \cite{ra01}, respectively. 
At zero temperature, the centroid energy of the ISGQR between 10 and 22 MeV is obtained at 14.6 MeV.
When compared with the experimental results of Ref. \cite{li10}, the ISGQR energy is overestimated around 1.7 MeV within our model calculations.
The first $2^{+}$ state energy is found at 1.3 MeV, in good agreement with the experimental data \cite{ra01}. 
We find that both the first $2^{+}$ state and the ISGQR exhibit strong collectivity with large strengths, as expected.
For instance, the first $2^{+}$ state is neutron dominated and takes contributions from neutron $3s_{1/2}\rightarrow 2d_{3/2}$ (23.9\%), $1g_{7/2}\rightarrow 2d_{3/2}$ (18.4\%), $1h_{11/2}\rightarrow 1h_{11/2}$ (15.2\%), $2d_{3/2}\rightarrow 2d_{3/2}$ (15.0\%), $1g_{7/2}\rightarrow 1g_{7/2}$ (6.3\%) and proton $1g_{9/2}\rightarrow 2d_{5/2}$ (3.8\%) excitations.
These results are also in agreement with previous theoretical studies \cite{ter06,scam13}. 
By increasing temperature, the ISGQR centroid energy decreases while the strength increases. 
The ISGQR centroid energies are obtained at 14.6, 14.2, and 14.1 MeV at $T=0$, $1$, and $2$ MeV, respectively. 
The low-energy part of the quadrupole spectrum is also impacted by the temperature. Similar to the $^{68}$Ni nucleus, the strength of the first $2^{+}$ state is quenched due to the phase transition of nuclei from superfluid to normal state.
Already at $T=1$ MeV, a new excited state is obtained at $E=4.75$ MeV with appreciable strength.
Apart from thermally unblocked proton states, thermal population of the neutron $1h_{11/2}$ state has an important impact on this low-energy strength. The most important contributions at $E=4.75$ MeV are coming from neutron $1h_{11/2}\rightarrow 2f_{7/2}$ (55.3\%), $1g_{9/2}\rightarrow 2d_{5/2}$ (7.1\%) and proton $1g_{9/2}\rightarrow 2d_{5/2}$ (19.5\%) transitions. The low-energy region below 4 MeV also starts to take contributions from thermally unblocked states at $T=1$ MeV. While the contribution of the excitations from thermally occupied states to the discretized continuum is rather small, the low-energy region is mainly impacted due to the changing occupation probabilities of states and contribution of new excitations around the Fermi level. By further increasing temperature, we obtain a small increase in the strength at around $E\approx2$ MeV due to the increasing diffuseness of the Fermi surface. These results are also in agreement with the FT-QRPA results of Ref. \cite{khan04}. The present study therefore provides a deep understanding of the underlying mechanism at the origin of the increase of the low-energy strength with temperature.
\begin{figure}[!ht]
  \begin{center}
\includegraphics[width=1\linewidth,clip=true]{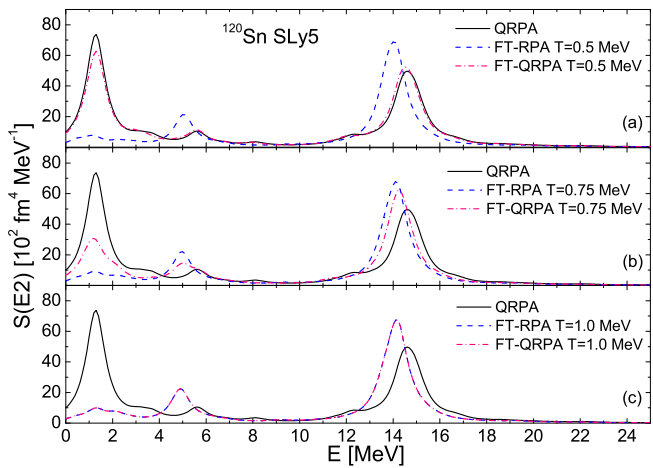}
  \end{center}
  \caption{The isoscalar quadrupole strength function in $^{120}$Sn calculated with QRPA ($T=0$ MeV), FT-RPA and FT-QRPA at $T=0.5$ (a), $0.75$ (b), and $1$ (c) MeV.} 
  \label{1202+2}
\end{figure}

Let us now investigate the importance of using an appropriate microscopic model in the calculation of the multipole response of the open-shell nuclei at finite temperatures. Since the isoscalar quadrupole response is rather sensitive to the temperature effects, it would be interesting to see the effect of the various microscopic models on the results. On this purpose, calculations are performed for the isoscalar quadrupole response in $^{120}$Sn
using QRPA ($T=0$ MeV), FT-RPA, and FT-QRPA. In panels (a), (b), and (c) of Fig. \ref{1202+2} we display the results for the isoscalar quadrupole response of $^{120}$Sn at $T=0.5$, $0.75$, and $1$ MeV, respectively. For $^{120}$Sn, the critical temperature is found at $T_{c}=0.83$ MeV. At $T=0.5$ MeV, it is clear that the FT-RPA and FT-QRPA results are quite different both in the high-energy and low-energy parts of the isoscalar quadrupole strength. 
Although we obtain a small increase in the isoscalar quadrupole strength below 4 MeV due to the opening of the new excitation channels, the FT-RPA predicts a small low-energy strength and underestimates the ISGQR energy when compared to the FT-QRPA results.
This difference is already indicating the important role of pairing correlations in open-shell nuclei.
At $T=0.75$ MeV, which is below the critical temperature, the FT-RPA and FT-QRPA results start to become compatible in the low-energy and high-energy parts of the strength due to the decrease of the pairing effects. Above the critical temperature, the pairing properties vanish and the phase change of nuclei from superfluid to the normal state occurs. Eventually, the FT-RPA and FT-QRPA give similar results at $T=1$ MeV, as expected. The results of the present study show that the self-consistent finite temperature QRPA is quite important in order to study and understand the temperature induced effects on the multipole response of open-shell nuclei below the critical temperatures. 

\section{Summary and Conclusions}
In this work, the fully self-consistent finite temperature QRPA has been used for the first time in order to study the effect of temperature on the evolution of the isovector dipole and isoscalar quadrupole responses in $^{68}$Ni and $^{120}$Sn nuclei.
The Skyrme-type SLy5 interaction is used in the calculations. 
The calculations were performed in two steps. First, the finite temperature HFBCS calculation was performed to determine the ground state properties of nuclei at finite temperatures. Then, the finite temperature QRPA was applied on top of it in order to study nuclear excited states ($J^{\pi}=1^{-}, 2^{+}$) in hot nuclei. 

At finite temperature, it is found that both
isovector giant dipole resonance energy and strength undergo minor changes.  
In addition, the contribution of additional configurations mainly impact the low-energy region of the dipole response in nuclei at finite temperatures. In general, temperature leads to the fragmentation and
distribution of the pygmy dipole resonance strength towards lower
energies due to the opening of the new excitation channels. We also find that the continuum plays an important role and the transitions from thermally unblocked states to the discretized continuum give rise to increase in the very low-energy region of the dipole strength at finite temperatures.
The effect of temperature on the low-energy region of the radiative dipole strength is also investigated in the $^{122}$Sn nucleus and compared with the experimental data. The results show that the very low-energy enhancement of the radiative dipole strength can be described by the effect of temperature on the dipole response.

The evolution of the isoscalar quadrupole response of
nuclei has also been analyzed at finite temperatures. While the ISGQR strength increases, the centroid energies decrease with increasing temperature. In the low-energy part of the isoscalar quadrupole spectrum,
the strength is considerably quenched in each nucleus above the critical temperature due to the disappearance of the pairing correlations. Furthermore, thermally occupied proton and neutron states give rise to the formation of the new excitation channels and increase the strength in the low-energy quadrupole spectrum due to the diffuse Fermi surface at finite temperatures.

By performing QRPA, FT-RPA and FT-QRPA
calculations, we have shown the importance of the use of the
appropriate microscopic models in the description of the nuclear
excitations in open-shell nuclei.
The results of the present study show that the
self-consistent finite temperature QRPA is relevant in order to study
and understand the temperature induced effects on the multipole
response of open-shell nuclei below the critical temperature.
As an extension of this work, it would be interesting to improve the FT-HFBCS formalism in order to remove the sharp phase transitions at critical temperatures. In addition, investigation of the spin-isospin resonances in open-shell nuclei within the FT-QRPA framework is also planned. It is also possible to extend the present FT-QRPA method in order to
study the Wigner-Seitz cells in neutron star crust at finite
temperatures with exact the continuum treatment. These issues may be the subject of forthcoming works.

\section{Acknowledgments}
The authors would like to thank Nguyen Van Giai for useful discussions and comments.
This work is supported by the Scientific and Technological Research Council of Turkey
(T\"{U}B\.{I}TAK) under project number MFAG-114F335, and it is partly supported by the National Natural Science Foundation of China under Grants No. 11305161. Funding from the European Union's Horizon 2020 research and innovation programme under grant agreement No 654002 is also acknowledged.

\appendix
\section{THE FINITE TEMPERATURE QRPA MATRICES}
\label{appendix}
In this appendix, we provide explicit forms of the finite temperature QRPA matrices in angular momentum coupled representation. 
The excitation operator in Eq. \ref{eq:ex} can be written in the coupled representation as
\begin{equation}
\begin{split}
\Gamma_{\nu}^{\dagger}=&\sum_{a\geq b} \Big\{X_{ab}^{\nu}A_{ab}^{\dagger}(JM) - Y_{ab}^{\nu}A_{ab}(\widetilde{JM}) \\
&+P_{ab}^{\nu}B_{ab}^{\dagger}(JM) - Q_{ab}^{\nu}B_{ab}(\widetilde{JM})\Big\}
\end{split}
\end{equation}
where $A_{ab}^{\dagger}(JM) [A_{ab}(\widetilde{JM})]$ and $B_{ab}^{\dagger}(JM) [B_{ab}(\widetilde{JM})]$ are the coupled 
two-quasiparticle creation-annihilation operators and one-quasiparticle creation plus one-quasiparticle destruction operators, respectively.
The coupled operators are given by
\begin{align}
\begin{split}
   A_{ab}^{\dagger}(JM)= N_{ab}(J)\sum_{m_{a}m_{b}}\left\langle j_{a}m_{a}j_{b}m_{b}\right|JM \rangle a_{a}^{\dagger}a_{b}^{\dagger},
	\end{split}\\
	\begin{split}
   A_{ab}(JM)= (A_{ab}^{\dagger}(JM))^{\dagger},
	\end{split}\\
	\begin{split}
   A_{ab}(\widetilde{JM})= (-1)^{J+M}A_{ab}(J-M),
	\end{split}\\
	\begin{split}
B_{ab}^{\dagger}(JM)= N_{ab}(J)\sum_{m_{a}m_{b}}(-1)^{j_{b}-m_{b}}\left\langle j_{a}m_{a}j_{b}-m_{b}\right|JM \rangle a_{a}^{\dagger}a_{b},
	\end{split}\\
	\begin{split}
B_{ab}(JM)=(B_{ab}^{\dagger}(JM))^{\dagger},
	\end{split}\\
	\begin{split}
   B_{ab}(\widetilde{JM})= (-1)^{J+M}B_{ab}(J-M)
	\end{split}
\end{align}
where $N_{ab}$ is the normalization factor and given as
\begin{equation}
 N_{ab}(J)=\frac{\sqrt{1+(-)^{J}\delta_{ab}}}{1+\delta_{ab}}.
\end{equation}
The finite temperature QRPA equations are obtained by the direct use of the equation of motion method (see Eq. \ref{eq:eom}).
In the quasiparticle representation, the nuclear Hamiltonian can be written as
\begin{equation}
\begin{split}
\label{eq:hm}
H=&H_{0}+\sum_{ij}H_{ij}^{11}a_{i}^{\dagger}a_{j}+\sum_{ijkl} H_{ijkl}^{22}a_{i}^{\dagger}a_{j}^{\dagger}a_{l}a_{k} \\
&+\sum_{ijkl}(H_{ijkl}^{40}a_{i}^{\dagger}a_{j}^{\dagger}a_{l}^{\dagger}a_{k}^{\dagger}+h.c.) \\
&+\sum_{ijkl}(H_{ijkl}^{31}a_{i}^{\dagger}a_{j}^{\dagger}a_{l}^{\dagger}a_{k}+h.c.)
\end{split}
\end{equation}
where $H_{0}$ is the mean-field Hamiltonian while $H_{11}$ represents the one-body creation-annihilation part.
In addition, the $H^{40}, H^{22}$ and $H^{31}$ parts contain two-body particle-hole (ph) and particle-particle (pp) matrix elements as well as $u$ and $v$ factors of the BCS equations. 
The explicit form of the quasiparticle representation of the Hamiltonian can be found in Refs. \cite{som83,suho07,ring80}. 
The finite temperature QRPA matrix equations are obtained from Eq. \ref{eq:eom} and given by
\begin{eqnarray}
\widetilde{A}_{abcd}=\left\langle  \left[A_{ab}(JM),\left[H,A_{cd}^{\dagger}(JM)\right]\right]\right\rangle, \\
\widetilde{B}_{abcd}=-\left\langle  \left[A_{ab}(JM),\left[H,A_{cd}(\widetilde{JM})\right]\right]\right\rangle, \\
\widetilde{C}_{abcd}=\left\langle \left[B_{ab}(JM),\left[H,B_{cd}^{\dagger}(JM)\right]\right]\right\rangle, \\
\widetilde{D}_{abcd}=-\left\langle \left[B_{ab}(JM),\left[H,B_{cd}(\widetilde{JM})\right]\right]\right\rangle, \\
\widetilde{a}_{abcd}=\left\langle \left[B_{ab}(JM),\left[H,A_{cd}^{\dagger}(JM)\right]\right]\right\rangle, \\
\widetilde{b}_{abcd}=-\left\langle \left[B_{ab}(JM),\left[H,A_{cd}(\widetilde{JM})\right]\right]\right\rangle,\\
\widetilde{a}_{abcd}^{+}=\left\langle \left[A_{ab}(JM),\left[H,B_{cd}^{\dagger}(JM)\right]\right]\right\rangle, \\
\widetilde{b}_{abcd}^{T}=-\left\langle \left[A_{ab}(JM),\left[H,B_{cd}(\widetilde{JM})\right]\right]\right\rangle.
\end{eqnarray}

The expectation values of the double commutators with respect to the $|BCS\rangle$ thermal vacuum give access to the finite temperature QRPA equations.
At finite temperatures, the operators read $\langle a_{j}^{\dagger}a_{i}\rangle=\delta_{ij}f_{i}$ and  
$\langle a_{j}a_{i}^{\dagger}\rangle=\delta_{ij}(1-f_{j})$. Consequently, the expectation values of the commutators of the coupled operators give
\begin{gather}
\begin{aligned}
&\langle BCS|[A_{ab}(JM),A_{cd}^{\dagger}(J'M')]|BCS\rangle \\ 
&=N_{ab}(J)^{2}\delta_{JJ'}\delta_{MM'}\left[\delta_{ac}\delta_{bd}-(-1)^{j_{a}+j_{b}+J}\delta_{ad}\delta_{bc}\right] \\
&\times(1-f_{a}-f_{b}),
\end{aligned} \\
\begin{aligned}
&\langle BCS|[B_{ab}(JM),B_{cd}^{\dagger}(J'M')]|BCS\rangle \\
&=N_{ab}(J)^{2}\delta_{JJ'}\delta_{MM'}\delta_{ac}\delta_{bd}(f_{b}-f_{a}).
\end{aligned}
\end{gather}

The explicit forms of the FT-QRPA matrices are obtained as
\begin{gather}
\begin{aligned}
A'_{abcd}&=(u_{a}u_{b}u_{c}u_{d}+v_{a}v_{b}v_{c}v_{d})V^{\text{pp}}_{abcd} \\
&+N_{ab}(J)N_{cd}(J)\big[(u_{a}v_{b}u_{c}v_{d}+v_{a}u_{b}v_{c}u_{d})V^{\text{ph}}_{a\bar{d}\bar{b}c} \\
&-(-1)^{j_{c}+j_{d}+J}(u_{a}v_{b}v_{c}u_{d}+v_{a}u_{b}u_{c}v_{d})V^{\text{ph}}_{a\bar{c}\bar{b}d}\big],
\end{aligned} \\
\begin{aligned}
B_{abcd}&=-(u_{a}u_{b}v_{c}v_{d}+v_{a}v_{b}u_{c}u_{d})V^{\text{pp}}_{ab\bar{c}\bar{d}} \\
&+N_{ab}(J)N_{cd}(J)\big[(u_{a}v_{b}v_{c}u_{d}+v_{a}u_{b}u_{c}v_{d})V^{\text{ph}}_{ad\bar{b}\bar{c}} \\
&-(-1)^{j_{c}+j_{d}+J}(u_{a}v_{b}u_{c}v_{d}+v_{a}u_{b}v_{c}u_{d})V^{\text{ph}}_{ac\bar{b}\bar{d}}\big],\\
\end{aligned} \\
\begin{aligned}
C'_{abcd}&=(u_{a}v_{b}u_{c}v_{d}+v_{a}u_{b}v_{c}u_{d})V^{\text{pp}}_{a\bar{b}c\bar{d}} \\
&+N_{ab}(J)N_{cd}(J)\big[(u_{a}u_{b}u_{c}u_{d}+v_{a}v_{b}v_{c}v_{d})V^{\text{ph}}_{adbc} \\
&+(-1)^{j_{c}+j_{d}+J}(u_{a}u_{b}v_{c}v_{d}+v_{a}v_{b}u_{c}u_{d})V^{\text{ph}}_{a\bar{c}b\bar{d}}\big],  
\end{aligned} \\
\begin{aligned} 
D_{abcd}&=(u_{a}v_{b}v_{c}u_{d}+v_{a}u_{b}u_{c}v_{d})V^{\text{pp}}_{a\bar{b}\bar{c}d} \\
&-N_{ab}(J)N_{cd}(J)\big[(u_{a}u_{b}v_{c}v_{d}+v_{a}v_{b}u_{c}u_{d})V^{\text{ph}}_{a\bar{d}b\bar{c}} \\
&+(-1)^{j_{c}+j_{d}+J}(u_{a}u_{b}u_{c}u_{d}+v_{a}v_{b}v_{c}v_{d})V^{\text{ph}}_{acbd}\big],
\end{aligned} \\
\begin{aligned}
a_{abcd}=&(v_{a}u_{b}v_{c}v_{d}-u_{a}v_{b}u_{c}u_{d})V^{\text{pp}}_{a\bar{b}cd} \\
&-N_{ab}(J)N_{cd}(J)\big[(v_{a}v_{b}v_{c}u_{d}-u_{a}u_{b}u_{c}v_{d})V^{\text{ph}}_{a\bar{d}bc}\\
&-(-1)^{j_{c}+j_{d}+J}(v_{a}v_{b}u_{c}v_{d}-u_{a}u_{b}v_{c}u_{d})V^{\text{ph}}_{a\bar{c}bd}\big],
\end{aligned} \\
\begin{aligned}
b_{abcd}=&-(v_{a}u_{b}u_{c}u_{d}-u_{a}v_{b}v_{c}v_{d})V^{\text{pp}}_{\bar{a}bcd} \\
&-N_{ab}(J)N_{cd}(J)\big[(v_{a}v_{b}u_{c}v_{d}-u_{a}u_{b}v_{c}u_{d})V^{\text{ph}}_{adb\bar{c}} \\
&-(-1)^{j_{c}+j_{d}+J}(v_{a}v_{b}v_{c}u_{d}-u_{a}u_{b}u_{c}v_{d})V^{\text{ph}}_{acb\bar{d}}\big],
\end{aligned} \\
\begin{aligned}
a_{abcd}^{+}=&(v_{a}v_{b}v_{c}u_{d}-u_{a}u_{b}u_{c}v_{d})V^{\text{pp}}_{abc\bar{d}} \\
&-N_{ab}(J)N_{cd}(J)\big[(v_{a}u_{b}v_{c}v_{d}-u_{a}v_{b}u_{c}u_{d})V^{\text{ph}}_{ad\bar{b}c}\\
&+(-1)^{j_{c}+j_{d}+J}(v_{a}u_{b}u_{c}u_{d}-u_{a}v_{b}v_{c}v_{d})V^{\text{ph}}_{\bar{a}cbd}\big],
\end{aligned} \\
\begin{aligned}
b_{abcd}^{T}=&(v_{a}v_{b}u_{c}v_{d}-u_{a}u_{b}v_{c}u_{d})V^{\text{pp}}_{ab\bar{c}d} \\
&+N_{ab}(J)N_{cd}(J)\big[(v_{a}u_{b}u_{c}u_{d}-u_{a}v_{b}v_{c}v_{d})V^{\text{ph}}_{\bar{a}dbc}\\
&+(-1)^{j_{c}+j_{d}+J}(v_{a}u_{b}v_{c}v_{d}-u_{a}v_{b}u_{c}u_{d})V^{\text{ph}}_{ac\bar{b}d}\big],
\end{aligned}
\end{gather}
where $V^{\text{ph}}$ and $V^{\text{pp}}$ represent the particle-hole and particle-particle effective interactions, respectively.
The effective interactions are defined as
$V^{\text{ph}}_{acbd}=\delta^{2}E(\rho,\kappa,\kappa^{*})/\delta\rho_{ba}\delta\rho_{dc}$,
$V^{\text{pp}}_{abcd}=\delta^{2}E(\rho,\kappa,\kappa^{*})/\delta\kappa^{*}_{ab}\delta\kappa_{cd}$
where the residual interaction contains both the spin-orbit and Coulomb interaction parts in order to achieve self-consistency.

%
\end{document}